\documentclass[12pt]{iopart}
\pdfoutput=1

\usepackage{graphicx}
\usepackage{iopams}

\newcommand{\pd}[2]{\frac{\partial #1}{\partial #2}}
\newcommand{\dd}[2]{\frac{\mathrm{d} #1}{\mathrm{d} #2}}
\newcommand{\bra}[1]{\left(#1\right)}
\newcommand{\sbra}[1]{\left[#1\right]}
\renewcommand{\epsilon}{\varepsilon}

\renewcommand{\leq}{\leqslant}

\renewcommand{\tilde}{\widetilde}
\newcommand{\coloneq}{\mathrel{\mathop:}=} 
\newcommand{\iint}{\int\!\!\!\int} 

\begin{document}
\title[Optimal control of plate shape with incompatible strain fields]{Optimal control of plate shape with incompatible strain fields}
\author{G.W.\ Jones$^{1,2}$ and L.\ Mahadevan$^2$}
\address{$^1$ School of Mathematics, University of Manchester, Oxford Road, Manchester, M13 9PL.}
\address{$^2$ School of Engineering and Applied Sciences. Harvard University, 29 Oxford Street, Cambridge, MA 02138; Wyss Institute for Biologically Inspired Engineering; Kavli Institute for Bionano Science and Technology}
\ead{\mailto{gareth.jones-10@manchester.ac.uk}, \mailto{lm@seas.harvard.edu}}
\begin{abstract}
A flat plate can bend into a curved surface if it experiences an inhomogeneous growth field. In this article a method is described that numerically determines the optimal growth field giving rise to an arbitrary target shape, optimizing for closeness to the target shape and for growth field smoothness. Numerical solutions are presented, for the full non-symmetric case as well as for simplified one-dimensional and axisymmetric geometries. This system can also be solved semi-analytically by positing an ansatz for the deformation and growth fields in a circular disk with given thickness profile. Paraboloidal, cylindrical and saddle-shaped target shapes are presented as examples, of which the last two exemplify a soft mode arising from a non-axisymmetric deformation of a structure with axisymmetric material properties.
\end{abstract}


\maketitle

\section{Background}\label{background}
Thin plate and shell-like structures are ubiquitous in nature, arising in such instances as leaves and petals in plants to heart valves and epithelial tissues in organisms.  They are also used in engineering and technological applications that range from flexible electronic circuits to  prosthetic tissue engineered valves valves, to large scale mechanical and civil structures.  A basic mathematical model that is used to describe and predict the mechanical behavior of these thin structures has its origins in elastic plate theory \cite{Love} having been well-studied for over a century. Most studies examine how the plate behaves in response to external and internal stimuli. These stimuli include not only forces applied to the surfaces and edges of the plate, but also more general effects such as thermal expansion, swelling, plastic deformation, and volumetric growth. While this forward problem remains a rich area of investigation in the mathematical, physical and engineering sciences, a natural question concerns the inverse problem of design --- how can we create optimal thin plate and shell-like structures for specific functions? Since shape is a precursor to function in many situations including the examples above, the simplest such inverse problem is that of asking how to shape a plate using boundary or bulk strains induced by external constraints or inhomogeneous growth. Here we examine just this inverse problem:  given a target shape that we want the plate to attain, how should the external or internal stimuli (henceforth ``control variables'') be chosen so that the plate is deformed into the target shape?

Early work on optimization of plate shape using boundary constraints includes studies focused on using normal traction on the plate surface to change its shape \cite{Bock-Hlavacek-Lovisek-1984,Bock-Hlavacek-Lovisek-1985,Gunzburger-Hou-1996}. However, while they were optimizing for a specific target configuration for the plate, their target was characterized by specifying both the normal displacement field $w$ and the Airy stress function $\chi$.  This formulation is unnecessarily  restrictive as a certain target \emph{shape} for the deformed plate can be provided by many combinations of $w$ and the in-plane displacements $v_1$, $v_2$ (which are linked to $\chi$) along the boundary. 

Recently, a new twist to this problem was added as a number of different groups have realized the ability to incorporate inelastic effects such as volumetric growth into elastic plate theory, a subject that has recently attracted much interest \cite{Sharon-Efrati-2010}. One area of particular interest is the imposition of inhomogeneous growth strains. These give rise to residual stresses which are relieved by the plate's buckling out of plane. This can be seen at the edges of certain leaves and flowers \cite{Liang-Mahadevan-2009,Liang-Mahadevan-2011}, which can have a rippled configuration due to inhomogeneous growth. Analogously, irreversible plastic deformation causes ripples at the edge of torn sheets of plastic. It is also possible to shape elastic plates made of gels and other polymeric materials that can swell by imbibing fluids \cite{Klein-Efrati-Sharon-2007,Sharon-Efrati-2010,Kim-Hanna-etal-2012,Kempaiah2014,Wei2014}. By blocking the ability of certain parts of the plate to swell or causing the plate to swell inhomogeneously, it is then possible to cause the plate to assume a variety of different shapes. In particular, these inhomogenous strains and boundary conditions cause the plate to deform, primarily by bending out of the plane, since that mode of deformation is usually inexpensive. That this is indeed possible in a controllable way was shown recently \cite{Dias-Hanna-Santangelo-2011} by analytically characterizing a class of in-plane volumetric growth that can transform an originally flat plate assuming certain symmetries in the shape, and  then validated experimentally.  

Here, we complement and generalize this idea to the case of using either boundary and bulk forcing and either in-plane or out-of-plane growth that can lead to variations in the natural curvature, all of which can be inhomogeneous. Our aim then is to find the growth strains so that this buckling --- and other growth-dependent deformations --- cause the reference plate shape to achieve a given target shape by balancing the requirements of closeness to the target while at the same time not having large inhomogeneities in the growth strain field (which are typically hard to engineer in technology or control in biology).

Our analysis will be more general than the specific instances outlined above in that we will develop a numerical method for arbitrary target shapes, and also consider not just in-plane growth but also active changes of curvature (caused by growth which is greater at one side of the plate than the other). The equations for growing plates are described in Section \ref{sect-growth-eqs}, and the optimization process (structurally similar to the work of Jones and Pereira \cite{Jones-Pereira-2014} that was started after this work was underway, but submitted earlier) is explained in Section \ref{sect-main-theory}. Following this we solve the system numerically for a general non-symmetric configuration (Section \ref{sect-2d}), and for simplified one-dimensional (Section \ref{sect-1d}) and axisymmetric (Section \ref{sect-axi}) geometries. Finally, in Section \ref{sect-simple} we use a semi-analytic approach on a circular disk to investigate how axisymmetric growth can give rise to so-called soft deformation modes.


\section{Theory}\label{sect-growth-eqs}

The equations governing volumetric growth in plates can be derived from one of two equivalent viewpoints:\ either by changing the definition of the reference metric, or by decomposing the strain tensors into growth and accommodation components. In the first approach, the reference metric of the plate is changed from its usual Euclidean form to a prescribed non-Euclidean metric. If a plate can be visualized as a collection of evenly-spaced points in a plane, connected by springs with a constant rest length, then imposing a non-Euclidean metric is equivalent to changing the spring rest lengths in such a way that a stress-free planar configuration of the points is impossible. This is kinematically equivalent to imposing an inhomogeneous in-plane growth field. Thus, a plate with an imposed non-Euclidean metric will tend to buckle out of plane in order to minimize its stored energy --- as long as the applied strains are sufficiently large. 
A second approach to this problem is to consider the elastic growth process directly, and to derive the equation in the limit of small strain and small plate thickness. 

This leads to very similar equations albeit approached from different perspectives --- from a differential geometric perspective \cite{Lewicka-Mahadevan-Pakzad-2011, Efrati-Sharon-Kupferman-2009-Elastic}, and formal  perturbation theory \cite{Dervaux-Ciarletta-BenAmar-2009}, \cite{Goriely-BenAmar-2007,Jones-Chapman-2012}, and bear deep similarities to the equations written down nearly half a century ago by Mansfield \cite{Mansfield-1962,Mansfield-1989} for the thermoelastic deformations for plates. In all cases, the nonlinear growth in an elastic body is kinematically described by a multiplicative decomposition of the deformation gradient  but in the plate limit the growth becomes an additive contribution to the strain fields.
In this section we will present the main equations, modified to account for varying plate thickness.

Growth is not the only phenomenon that can be described using this formalism; both thermoelastic expansion and plastic deformation are also kinematically described (especially in the small-thickness limit of plate theory) by additive decompositions of the strain tensors. The difference between these three theories is of course in how the non-elastic part (growth, thermal expansion, plastic deformation) is described, and in how these effects alter the properties of the material (including material density, stiffness tensors, and porosity). We will assume that the non-elastic parts of the strain tensor are small, so that these higher-order effects can be neglected.

With this in mind, we define a plate using Cartesian coordinates $\boldsymbol{X}=(X_1,X_2)$, with its deformation characterized by the in-plane displacements $v_\alpha(\boldsymbol{X})$ and out-of-plane displacement $w(\boldsymbol{X})$, where Greek indices vary over $1,2$. The growth in the plate may be characterized by the growth strains $\Gamma_{\alpha\beta}$ and $\Psi_{\alpha\beta}$, such that the in-plane strain $\gamma_{\alpha\beta}$ and the change-of-curvature tensor $\rho_{\alpha\beta}$ may be additively decomposed into growth and accommodation components: $\gamma_{\alpha\beta} = \Gamma_{\alpha\beta} + \gamma_{\alpha\beta}^\mathrm{e}$ and $\rho_{\alpha\beta} = \Psi_{\alpha\beta} + \rho_{\alpha\beta}^\mathrm{e}$. This decomposition is valid if the strain fields remain small. In terms of displacement the elastic accommodation strain tensors are thus given by
\begin{eqnarray}
\gamma_{\alpha\beta}^\mathrm{e}=\frac{1}{2}(v_{\alpha,\beta}+v_{\beta,\alpha})+\frac{1}{2}w_{,\alpha}w_{,\beta}-\Gamma_{\alpha\beta},\label{ga-e}\\
\rho_{\alpha\beta}^\mathrm{e} = w_{,\alpha\beta} - \Psi_{\alpha\beta},\label{rh-e}
\end{eqnarray}
and an index preceded by a comma indicates differentiation with respect to that coordinate. The elastic energy density is given by
\begin{equation}
W^\mathrm{el}=\frac{1}{2} DA_{\alpha\beta\lambda\mu} \gamma_{\alpha\beta}^\mathrm{e}\gamma_{\lambda\mu}^\mathrm{e} + \frac{1}{2} BA_{\alpha\beta\lambda\mu} \rho_{\alpha\beta}^\mathrm{e}\rho_{\lambda\mu}^\mathrm{e}
\end{equation}
(applying the summation convention), where
\begin{eqnarray}
D=\frac{Eh}{1-\nu^2},\qquad B=\frac{Eh^3}{12(1-\nu^2)},\\
A_{\alpha\beta\lambda\mu}=\bra{\frac{1-\nu}{2}} (\delta_{\alpha\lambda}\delta_{\beta\mu} + \delta_{\alpha\mu}\delta_{\beta\lambda}) + \nu\delta_{\alpha\beta}\delta_{\lambda\mu},\label{A}
\end{eqnarray}
and $E$, $\nu$ and $h$ are the Young's modulus, Poisson ratio and thickness of the plate respectively. We scale the displacements $v$ and $w$ with $L$, a typical lengthscale of the problem; $\Psi_{\alpha\beta}$ with $1/L$; the variable thickness $h$ with typical value $h_0$, leading to typical values $D_0=Eh_0/(1-\nu^2)$, $B_0=Eh_0^3/12(1-\nu^2)$ for the stiffnesses. Finally we define $\beta=B_0/(D_0L^2)=h_0^2/(12L^2)$ to be the dimensionless stiffness ratio.

The dimensionless equations governing the plate deformation under the action of the growth strains $\Gamma_{\alpha\beta}$ and $\Psi_{\alpha\beta}$ (assuming no surface loading) are the generalized F\"oppl--von K\'arm\'an (FvK) equations:
\begin{eqnarray}
\frac{1}{(1-\nu^2)}\nabla^2\bra{\frac{1}{h}\nabla^2\chi}-\frac{1}{(1-\nu)}\sbra{\frac{1}{h},\chi}+\frac{1}{2}[w,w]+\lambda^\mathrm{g}=0,\label{fvk1}\\
\beta\nabla^2\bra{h^3\nabla^2w}-\beta(1-\nu)[h^3,w]-[w,\chi]+\beta\Phi^\mathrm{g}=0.\label{fvk2}
\end{eqnarray}
In these expressions, $[f,g]=f_{,11}g_{,22}-2f_{,12}g_{,12}+f_{,22}g_{,11}$, and $\chi$ is the Airy stress function for the plate, defined (non-dimensionally) through the stress resultant tensor,
\begin{equation}
N_{\alpha\beta}\coloneq hA_{\alpha\beta\lambda\mu}\gamma_{\lambda\mu}^\mathrm{e}= \epsilon_{\alpha\lambda}\epsilon_{\beta\mu}\chi_{,\lambda\mu},\label{chi-definition}
\end{equation}
where
\begin{equation}
\epsilon_{\alpha\beta}=\textstyle\bra{\textstyle\begin{array}{cc}0&1\\-1&0\end{array}}
\end{equation}
is the two-dimensional alternating tensor. The source terms in (\ref{fvk1})--(\ref{fvk2}) due to growth are
\begin{eqnarray}
\lambda^\mathrm{g} = \epsilon_{\alpha\beta}\epsilon_{\lambda\mu} \Gamma_{\alpha\lambda,\beta\mu} = \Gamma_{11,22} + \Gamma_{22,11} - 2\Gamma_{12,12},
\end{eqnarray}
and
\begin{equation}
\Phi^\mathrm{g} = -(1-\nu)\partial_{\alpha\beta}(h^3\Psi_{\alpha\beta}) - \nu\partial_{\beta\beta}(h^3\Psi_{\alpha\alpha})
\end{equation}
or
\begin{eqnarray}
\fl\Phi^\mathrm{g}= -\left[(h^3\Psi_{11})_{,11} + \nu(h^3\Psi_{11})_{,22} + \nu(h^3\Psi_{22})_{,11}\right.\\
+ \left.(h^3\Psi_{22})_{,22} + 2(1-\nu)(h^3\Psi_{12})_{,12}\right].
\end{eqnarray}

Note that for isotropic growth, \emph{i.e.}\ $\Gamma_{\alpha\beta}=\Gamma\delta_{\alpha\beta}$ and $\Psi_{\alpha\beta}=\Psi\delta_{\alpha\beta}$, the source terms simplify to $\lambda^\mathrm{g}=\nabla^2\Gamma$ and $\Phi^\mathrm{g}=-(1+\nu)\nabla^2(h^3\Psi)$. Furthermore, if the material was perfectly accommodating of growth, the equations simplify to
\begin{equation}
\frac{\kappa_\mathrm{G}}{2}+\lambda_\mathrm{g}=0,\qquad[w,\chi]=0,
\end{equation}
where $\kappa_\mathrm{G}=[w,w]$ is the Gaussian curvature of the deformed surface.

The FvK equations are solved with appropriate boundary conditions. If $t_\alpha$, $n_\alpha$ are the components of the tangent and normal vectors to the plate edge, then the natural boundary conditions, corresponding to force-free and moment-free conditions, are
\begin{eqnarray}
N_{\alpha\beta}n_\beta=0 \qquad \Rightarrow\qquad\chi=0\quad \textnormal{and}\quad\partial_n\chi=0,\label{free-bcs-1}\\
M_{\alpha\beta,\alpha}n_\beta+\partial_t(M_{\alpha\beta}n_\alpha t_\beta)=0,\label{free-bcs-2}\\
M_{\alpha\beta}n_\alpha n_\beta=0\label{free-bcs-3},
\end{eqnarray}
where $N_{\alpha\beta}$ is the stress resultant tensor given in (\ref{chi-definition}), and $M_{\alpha\beta}$ is the (dimensionless) moment resultant tensor, $M_{\alpha\beta}=\beta h^3A_{\alpha\beta\lambda\mu}\rho_{\lambda\mu}^\mathrm{e}$. We will also be applying pinned boundary conditions, for which (\ref{free-bcs-1})--(\ref{free-bcs-2}) are replaced by $w=0$, $v_\alpha=0$.

The F\"oppl--von K\'arm\'an equations (\ref{fvk1})--(\ref{fvk2}) outlined above do not involve the tangential displacement field $v_1$, $v_2$ directly. Thus an extra step would be needed to calculate $v_1$ and $v_2$ from $\chi$ before measuring the distance between the deformed plate and the target shape. An alternative to this approach is to write the system explicitly in terms of the three displacement components $v_1$, $v_2$, $w$.

We find it more convenient to write these equations in weak form, as they may be solved straightforwardly using finite elements. The F\"oppl--von K\'arm\'an equations with growth were written in weak form by Lewicka \etal \cite{Lewicka-Mahadevan-Pakzad-2011}. However, in their formulation the normal displacements $w$ are required to be twice differentiable. As we will be using linear finite elements, we modify the equations following Reinhart \cite{Reinhart-1982}, who treats the curvature $\rho_{\alpha\beta}$ as three new independent variables, with three additional weak-form equations to solve.

In summary, the six equations to solve for the six variables $v_1$, $v_2$, $w$, $\rho_{11}$, $\rho_{12}$, $\rho_{22}$ are shown below. Quantities with a tilde are the variations; the weak equations hold for all admissible (once-differentiable) values of these variations.
\begin{eqnarray}
\fl\iint_\Omega\sbra{\pd{\tilde{v}_1}{X}N_{11}+\pd{\tilde{v}_1}{Y}N_{12}-\tilde{v}_1q_1}\,\mathrm{d}^2\boldsymbol{X}=0,\label{weakform1}\\
\fl\iint_\Omega\sbra{\pd{\tilde{v}_2}{X}N_{12}+\pd{\tilde{v}_2}{Y}N_{22}-\tilde{v}_2q_2}\,\mathrm{d}^2\boldsymbol{X}=0,\\
\fl\iint_\Omega\sbra{\tilde{\rho}_{11}(\rho_{11}+\nu\rho_{22})+\pd{\tilde{\rho}_{11}}{X}\pd{w}{X}+\nu\pd{\tilde{\rho}_{11}}{Y}\pd{w}{Y}}\,\mathrm{d}^2\boldsymbol{X}\nonumber\\
 = \oint_{\partial\Omega}\tilde{\rho}_{11}\bra{\pd{w}{X}n_1+\nu\pd{w}{Y}n_2}\,\mathrm{d}s,\\
\fl\iint_\Omega\sbra{2\tilde{\rho}_{12}\rho_{12}+\pd{\tilde{\rho}_{12}}{X}\pd{w}{Y}+\pd{\tilde{\rho}_{12}}{Y}\pd{w}{X}}\,\mathrm{d}^2\boldsymbol{X} = \oint_{\partial\Omega}\tilde{\rho}_{12}\bra{\pd{w}{X}n_2+\pd{w}{Y}n_1}\,\mathrm{d}s,\\
\fl\iint_\Omega\sbra{\tilde{\rho}_{22}(\nu\rho_{11}+\rho_{22})+\nu\pd{\tilde{\rho}_{22}}{X}\pd{w}{X}+\pd{\tilde{\rho}_{22}}{Y}\pd{w}{Y}}\,\mathrm{d}^2\boldsymbol{X}\nonumber\\
 = \oint_{\partial\Omega}\tilde{\rho}_{22}\bra{\nu\pd{w}{X}n_1+\pd{w}{Y}n_2}\,\mathrm{d}s,\\
\fl\iint_\Omega\left[p\tilde{w}+\pd{\tilde{w}}{X}\bra{-\pd{w}{X}N_{11}-\pd{w}{Y}N_{12}+\pd{M_{11}}{X}+\pd{M_{12}}{Y}}\right.\nonumber\\
+\left.\pd{\tilde{w}}{Y}\bra{-\pd{w}{X}N_{12}-\pd{w}{Y}N_{22}+\pd{M_{12}}{X}+\pd{M_{22}}{Y}}\right]\,\mathrm{d}^2\boldsymbol{X}\nonumber\\
=\oint_{\partial\Omega}\sbra{\pd{\tilde{w}}{X}(M_{11}n_1+M_{12}n_2)+\pd{\tilde{w}}{Y}(M_{12}n_1+M_{22}n_2)}\,\mathrm{d}s.\label{weakform6}
\end{eqnarray}

In these expressions,
\begin{eqnarray}
h^{-1}N_{11}=\pd{v_1}{X}+\nu\pd{v_2}{Y}+\frac{1}{2}\bra{\pd{w}{X}}^2+\frac{\nu}{2}\bra{\pd{w}{Y}}^2-(\Gamma_{11}+\nu\Gamma_{22}),\\
h^{-1}N_{12}=\frac{(1-\nu)}{2}\bra{\pd{v_1}{Y}+\pd{v_2}{X}+\pd{w}{X}\pd{w}{Y}-2\Gamma_{12}},\\
h^{-1}N_{22}=\nu\pd{v_1}{X}+\pd{v_2}{Y}+\frac{\nu}{2}\bra{\pd{w}{X}}^2+\frac{1}{2}\bra{\pd{w}{Y}}^2-(\nu\Gamma_{11}+\Gamma_{22}),\\
M_{11}=\beta h^3\sbra{\rho_{11}-\Psi_{11}+\nu(\rho_{22}-\Psi_{22})},\\
M_{12}=\beta h^3(1-\nu)(\rho_{12}-\Psi_{12}),\\
M_{22}=\beta h^3\sbra{\nu(\rho_{11}-\Psi_{11})+\rho_{22}-\Psi_{22}},\label{M22}
\end{eqnarray}
and we have included the normal and tangential surface tractions, $p$ and $q_\alpha$ respectively, for completeness. $\Omega$ is the domain of the undeformed plate.

On solving equations (\ref{weakform1})--(\ref{weakform6}) in the space of once-differentiable functions, the natural boundary conditions are the free boundary conditions (\ref{free-bcs-1})--(\ref{free-bcs-3}). For pinned boundary conditions the space of admissible functions must in addition specify that $v_1=v_2=w=\tilde{v}_1=\tilde{v}_2=\tilde{w}=0$ on the plate boundary.

For clamped boundary conditions (for instance, in the example provided in the introduction) the right-hand sides of all six equations must be set to zero, in order to impose $\boldsymbol{\nabla}w=\boldsymbol{0}$ on the boundary without specifying (\ref{free-bcs-3}) there also.

\section{Optimal control}\label{sect-main-theory}
While the previous section allows one to calculate the plate displacements subject to certain stimuli (growth fields, surface tractions, edge displacements), the key calculation from our viewpoint is to find what form of stimulus will give a desired property of the displacement field. Abstractly, we denote the stimuli as control variables $\boldsymbol{d}$ and the displacement and curvatures as state variables $\boldsymbol{u}$. Then the condition on the displacement field can be written as a minimization of a certain functional $\mathcal{E}_\mathrm{D}(\boldsymbol{u})$ of the state variables. Thus we obtain a PDE-constrained optimization problem:
\begin{eqnarray}
\min_{\boldsymbol{u},\boldsymbol{d}}\mathcal{E}_\mathrm{D}(\boldsymbol{u})\qquad\textrm{subject to }c_i(\boldsymbol{u},\boldsymbol{d})=0\label{PDEconopt}
\end{eqnarray}
for $i=1,\ldots,m$. The equations $c_i=0$ are the constraints, which comprise the FvK equations (\ref{weakform1})--(\ref{weakform6}).

The problem (\ref{PDEconopt}) is ill-posed, since there will be many combinations of $\boldsymbol{u}$ and $\boldsymbol{d}$ that minimize $\mathcal{E}_\mathrm{D}$, and non-smooth solutions are often the most accessible to numerical methods. Thus a regularization term $\mathcal{E}_\mathrm{S}$ must be added to $\mathcal{E}_\mathrm{D}$, so that some property of the control variables is minimized. Tikhonov regularization is a commonly-encountered example of this method. The optimization problem becomes
\begin{eqnarray}
\min_{\boldsymbol{u},\boldsymbol{d}}\sbra{ \mathcal{E}_\mathrm{D}(\boldsymbol{u})+\eta\mathcal{E}_\mathrm{S}(\boldsymbol{d}) }\qquad\textrm{subject to }c_i(\boldsymbol{u},\boldsymbol{d})=0
\end{eqnarray}
for $i=1,\ldots,m$. The parameter $\eta$ is chosen as a trade-off between numerical well-posedness and adherence to the requirement that the target displacement be met.

\subsection{Application to edge-displacement problems}
As an example of the situation that we envisage, consider a flat plate of arbitrary shape. The edges of the plate are clamped and are allowed to be displaced in-plane. In this situation the inverse problem to be solved is how to choose these edge displacements so that the interior is deformed into a given configuration. For example, consider a circular plate of radius $1$. How should the clamped edges be deformed so that the center point of the plate attains a given vertical displacement, $w|_{r=0}=w^*$?

The theory of section \ref{sect-main-theory} can be used to solve this problem --- and other plate optimization problems --- with certain modifications. The constraints to the problem are the FvK equations (\ref{weakform1})--(\ref{weakform6}), with $\Gamma=\Psi=0$ and $p=q_\alpha=0$. The boundary conditions are clamped, so the right-hand sides of (\ref{weakform1})--(\ref{weakform6}) are set to zero. At the boundary we impose $v_i=v_i^\mathrm{c}$ and $w=0$ (and $\tilde{v}_i=\tilde{w}=0$), where $v_i^\mathrm{c}$ are the prescribed edge displacements, used as control variables.

Finally we must specify an objective function. An appropriate form is $\mathcal{E}_\mathrm{D}=(w|_{r=0}-w^*)^2$ --- but as we have seen, the problem is ill-posed without a regularization term $\mathcal{E}_\mathrm{S}$. For this problem we set
\begin{equation}
\mathcal{E}_\mathrm{S}=\oint_{\partial\Omega}\sbra{\bra{\pd{v_1}{s}}^2+\bra{\pd{v_2}{s}}^2}\mathrm{d}s.
\end{equation}
Then the problem is solved by minimizing $\mathcal{E}_\mathrm{D}+\eta\mathcal{E}_\mathrm{S}$ --- subject to the FvK equations with zero growth and clamped boundary conditions --- by varying the state variables $w$, $v_i$ and the control variables $v_i^\mathrm{c}$.

\begin{figure}[ht]
\centering
\includegraphics[clip,width=0.7\textwidth]{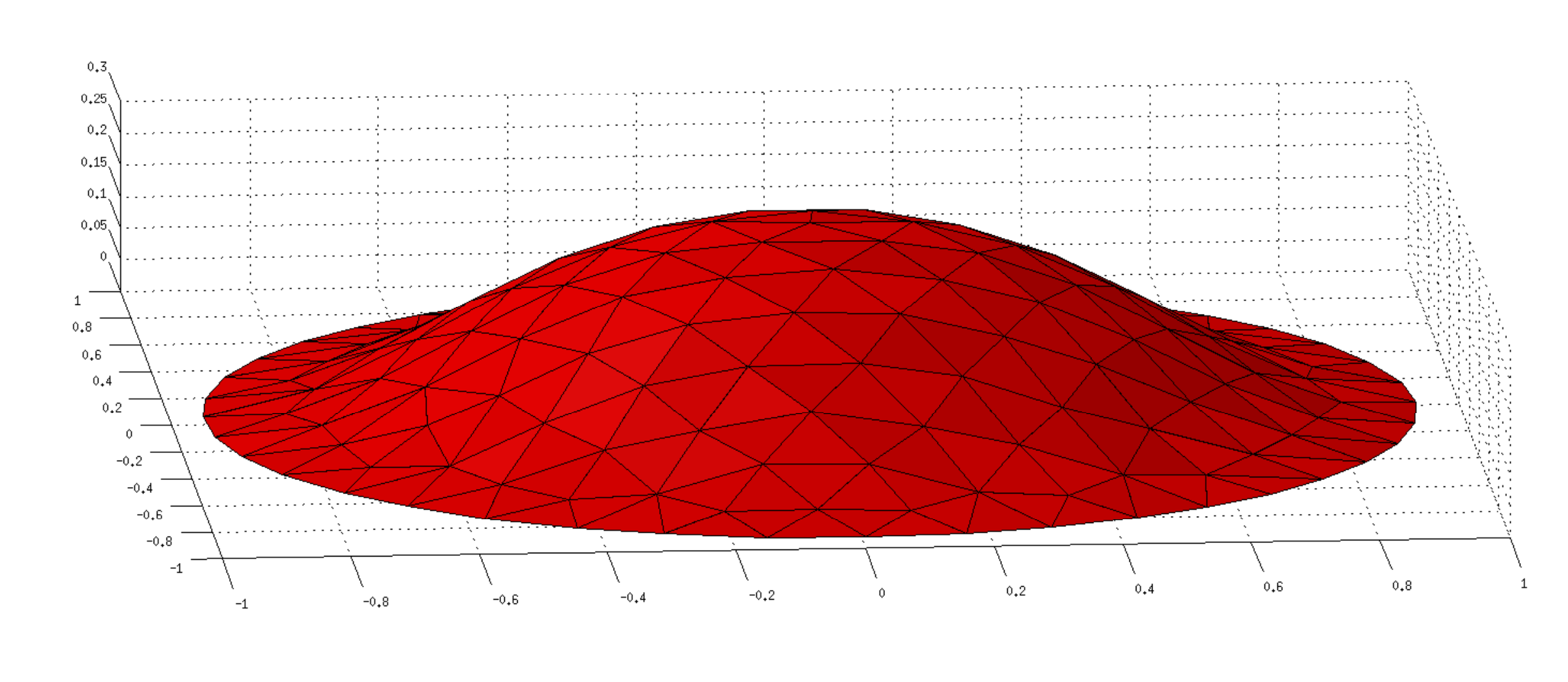}
\caption{The output of the example optimization procedure, with a given vertical displacement $w^*=0.3$ at the plate center achieved by solving for the plate edge displacements. Plate thickness $h=0.11$, Poisson ratio $\nu=0.3$ and regularization parameter $\eta=0.01$.}
\label{circle_bump}
\end{figure}

In Figure \ref{circle_bump} we display the results of this optimization calculation for $w^*=0.3$. 

\subsection{Application to growing plates}\label{sect-opt-growth}
We will now formulate an optimization problem for the growing plate --- in other words, to determine the optimal growth strains that allow the plate to achieve a given target shape. We propose that the optimal solution should minimize the functional $\mathcal{E}=\mathcal{E}_\mathrm{D}+\mathcal{E}_\mathrm{S}$, with the regularization parameters to be introduced later. In this functional, $\mathcal{E}_\mathrm{D}$ is a measure of the distance between the deformed plate and the target shape and $\mathcal{E}_\mathrm{S}$ is a regularization term which has the effect of smoothing the growth fields. In general the solution will therefore comprise a balance between closeness to the target shape, and spatial smoothness of the growth fields.

The Fr\'echet distance \cite{Alt-Buchin-2010} and Hausdorff distance \cite{Aspert-SantaCruz-Ebrahimi-2002} are general measures of the distance between two surfaces in three dimensions. However, if the target is known as an analytic function, simpler formulations are possible.

If the target shape and plate deformations are axisymmetric or otherwise one-dimensional, we can make use of the following scaled arclength implementation. Consider a 1D plate of length $1$, with a target shape $z=f(x)$ for $x\in(0,x_\mathrm{max})$. Then the parametric definitions of the curves traced out by the deformed plate (under in-plane displacement $v$ and normal displacement $w$) and the target shape are respectively
\begin{eqnarray}
\big(X+v(X),w(X)\big)& \qquad\textrm{for }X\in(0,1),\\
\big(x,f(x)\big)& \qquad\textrm{for }x\in(0,x_\mathrm{max}).
\end{eqnarray}
The arclengths of the curves are then
\begin{eqnarray}
S(X)&=\int_0^X\sbra{(1+v'(\bar{X}))^2+w'(\bar{X})^2}^{1/2} \mathrm{d}\bar{X},&\qquad S_\mathrm{max}=S(1),\label{eq:SX}\\
s(x)&=\int_0^x\sbra{1+f'(\bar{x})^2}^{1/2}\mathrm{d}\bar{x},&\qquad s_\mathrm{max}=s(x_\mathrm{max})\label{eq:sx}
\end{eqnarray}
respectively. These can be inverted to give $\tilde{X}(S)$, $\tilde{x}(s)$, and thus the deformed and target shapes parametrized by arclength:
\begin{eqnarray}
\bra{\tilde{X}(S)+v(\tilde{X}(S)),w(\tilde{X}(S))}&\qquad \textrm{for }S\in(0,S_\mathrm{max}),\\
\bra{\tilde{x}(s),f(\tilde{x}(s))}&\qquad \textrm{for }s\in(0,s_\mathrm{max}).
\end{eqnarray}
We can then define a distance function $\mathcal{D}$ by scaling $S$ and $s$ to provide a correspondence between these two parametrizations:\ let
\begin{eqnarray}
\fl\mathcal{D}^\mathrm{arc}(\sigma)^2=\sbra{\tilde{X}(S_\mathrm{max}\sigma)+v(\tilde{X}(S_\mathrm{max}\sigma))-\tilde{x}(s_\mathrm{max}\sigma)}^2\nonumber\\
 + \sbra{w(\tilde{X}(S_\mathrm{max}\sigma))-f(\tilde{x}(s_\mathrm{max}\sigma))}^2\qquad\textrm{for }\sigma\in(0,1),\label{arclength-distance}
\end{eqnarray}
and define
\begin{equation}
\mathcal{E}^\mathrm{arc}_\mathrm{D} =\eta_\mathrm{D}\int_0^1\mathcal{D}^\mathrm{arc}(\sigma)^2\mathrm{d}\sigma
\end{equation}
for some tunable parameter $\eta_\mathrm{D}$.

In the more general two-dimensional case, if the target shape is given as an elevation --- \emph{i.e.}\ $z=f(x,y)$ in Eulerian components --- then we may write the distance between the deformed plate and the target as
\begin{equation}
\mathcal{D}=w(X,Y)-f(X+v_1(X,Y),Y+v_2(X,Y)),
\end{equation}
and minimize $\eta_\mathrm{D}\iint_\Omega\mathcal{D}^2\,\mathrm{d}^2\boldsymbol{X}$. Note however that we must impose an additional constraint that the boundary of the undeformed plate must be mapped to the boundary of the target shape. This can be achieved by adding a term $\mathcal{E}_\mathrm{E}$ to the objective function, which is a measure of the distance between these two boundaries and can be calculated using the arclength method described above. Specifically, if
\begin{eqnarray}
(X_\mathrm{b}(\theta),Y_\mathrm{b}(\theta),0),\\
(x_\mathrm{b}^\mathrm{d}(\theta),y_\mathrm{b}^\mathrm{d}(\theta),z_\mathrm{b}^\mathrm{d}(\theta))\coloneq (X_\mathrm{b}+v_1(X_\mathrm{b},Y_\mathrm{b}), Y_\mathrm{b}+v_2(X_\mathrm{b},Y_\mathrm{b}), w(X_\mathrm{b},Y_\mathrm{b})),\\
(x_\mathrm{b}(\theta),y_\mathrm{b}(\theta),z_\mathrm{b}(\theta))
\end{eqnarray}
are the parametric representations of the undeformed, deformed, and target boundaries respectively, then by analogy with (\ref{eq:sx}),
\begin{eqnarray}
S(\theta)&=\int_0^\theta({x_\mathrm{b}^\mathrm{d}}'(\bar{\theta})^2+ {y_\mathrm{b}^\mathrm{d}}'(\bar{\theta})^2+ {z_\mathrm{b}^\mathrm{d}}'(\bar{\theta})^2)^{1/2}\,\mathrm{d}\bar{\theta}, &\qquad S_\mathrm{max}=S(2\pi),\\
s(\theta)&=\int_0^\theta(x_\mathrm{b}'(\bar{\theta})^2+ y_\mathrm{b}'(\bar{\theta})^2+ z_\mathrm{b}'(\bar{\theta})^2)^{1/2}\,\mathrm{d}\bar{\theta}, &\qquad s_\mathrm{max}=s(2\pi).
\end{eqnarray}
Invert these to give $\theta_S(S)$ and $\theta_s(s)$, and thus the deformed and target shapes parametrized by arclength:
\begin{eqnarray}
(x_\mathrm{b}^\mathrm{d}(\theta_S(S)),y_\mathrm{b}^\mathrm{d}(\theta_S(S)),z_\mathrm{b}^\mathrm{d}(\theta_S(S))),\\
(x_\mathrm{b}(\theta_s(s)),y_\mathrm{b}(\theta_s(s)),z_\mathrm{b}(\theta_s(s))).\label{xbybzb}
\end{eqnarray}
Then
\begin{eqnarray}
\fl\mathcal{E}_\mathrm{E}=\eta_\mathrm{E}\int_0^1\left\{ \sbra{x_\mathrm{b}^\mathrm{d}(\theta_S(S_\mathrm{max}\sigma))-x_\mathrm{b}(\theta_s(s_\mathrm{max}\sigma))}^2+ \sbra{y_\mathrm{b}^\mathrm{d}(\theta_S(S_\mathrm{max}\sigma))-y_\mathrm{b}(\theta_s(s_\mathrm{max}\sigma))}^2\right.\nonumber\\
\left.
+\sbra{z_\mathrm{b}^\mathrm{d}(\theta_S(S_\mathrm{max}\sigma))-z_\mathrm{b}(\theta_s(s_\mathrm{max}\sigma))}^2\right\}\mathrm{d}\sigma,\label{EE}
\end{eqnarray}
and $\mathcal{E}_\mathrm{D}=\eta_\mathrm{D}\iint_\Omega\mathcal{D}^2\,\mathrm{d}^2\boldsymbol{X}+\mathcal{E}_\mathrm{E}$.

The regularization term $\mathcal{E}_\mathrm{S}$ noted earlier is given by
\begin{equation}
\mathcal{E}_\mathrm{S}=\iint_{\Omega}\sbra{\frac{\eta_\Gamma}{2} \Gamma_{\alpha\beta,\gamma}\Gamma_{\alpha\beta,\gamma} + \frac{\eta_\Psi}{2} \Psi_{\alpha\beta,\gamma}\Psi_{\alpha\beta,\gamma}}\,\mathrm{d}^2\boldsymbol{X},
\end{equation}
where $\eta_\Gamma$ and $\eta_\Psi$ are tunable parameters. This objective function embodies the principle that the gradients of the growth strains in the optimal solution should be as small as possible. (For isotropic growth the regularizing term becomes $\eta_\Gamma|\boldsymbol{\nabla}\Gamma|^2+\eta_\Psi|\boldsymbol{\nabla}\Psi|^2$.) One practical reason for this restriction on the growth strains is that if we were to experimentally verify the solutions obtained by the optimization process, we would want the solution to be as insensitive as possible to manufacturing errors, which would be hard to achieve if $\Gamma_{\alpha\beta}$ and $\Psi_{\alpha\beta}$ varied rapidly across the undeformed plate.

The minimization of $\mathcal{E}$ will be subject to the constraint that the control variables $\Gamma_{\alpha\beta}$, $\Psi_{\alpha\beta}$ and state variables $\rho_{\alpha\beta}$, $w$, and $v_\alpha$ satisfy the modified F\"oppl--von K\'arm\'an equations (\ref{weakform1})--(\ref{weakform6}), in the appropriate function spaces (surface tractions $p$, $q_\alpha$ are set to zero).

Thus the optimization problem can be stated as follows:
\begin{eqnarray}
\min_{\Gamma_{\alpha\beta},\Psi_{\alpha\beta},v_\alpha,w,\rho_{\alpha\beta}}\sbra{\mathcal{E}_\mathrm{D}+ \iint_{\Omega}\bra{\frac{\eta_\Gamma}{2} \Gamma_{\alpha\beta,\gamma}\Gamma_{\alpha\beta,\gamma} + \frac{\eta_\Psi}{2} \Psi_{\alpha\beta,\gamma}\Psi_{\alpha\beta,\gamma}}\,\mathrm{d}^2\boldsymbol{X}}\label{nondim-objective}\\
\textrm{subject to the FvK equations (\ref{weakform1})--(\ref{weakform6}).}\nonumber
\end{eqnarray}

In Section \ref{sect-2d} we will outline some numerical solutions of the optimization problem (\ref{nondim-objective}), first in its full two-dimensional implementation, followed by simplified one-dimensional situations, namely a beam and an axisymmetric target shape. Following this we will discuss a semi-analytic approach, where growth leading to simple target shapes can give rise to soft deformation modes.

\section{Results: full two-dimensional shapes}\label{sect-2d}
To solve (\ref{nondim-objective}), we need to discretize the variables. To this end, the space of admissible solutions to (\ref{weakform1})--(\ref{weakform6}) is approximated by the space of piecewise affine functions, and the domain $\Omega$ is triangulated (for our calculations we used the DistMesh routine \cite{Persson-Strang-2004}). To simplify calculations in this section, the domain $\Omega$ is a circle of radius $1$, the thickness $h$ is set to $1$ and growth is isotropic ($\Gamma_{\alpha\beta} = \Gamma\delta_{\alpha\beta}$, $\Psi_{\alpha\beta} = \Psi\delta_{\alpha\beta}$). The control variables $\Gamma$, $\Psi$ and the state variables $w$, $v_1$, $v_2$, $\rho_{11}$, $\rho_{12}$, $\rho_{22}$ are all set to be piecewise affine over each triangle element, so that the function values at each node of the triangulation become the discrete variables to be solved for, as in the standard linear finite element approach. We used the sparse SQP solver \verb|e04vh| of the NAG toolbox, based on the software package SNOPT \cite{Gill-Murray-Saunders-2005}. This algorithm is well suited to such discrete numerical nonlinear optimization problems, and may be accessed through an interface to the numerical analysis package MATLAB. For a more thorough overview of the numerical procedure, refer to \ref{proc-appendix}.

\begin{figure}[ht]
\centering
\includegraphics[width=1\textwidth]{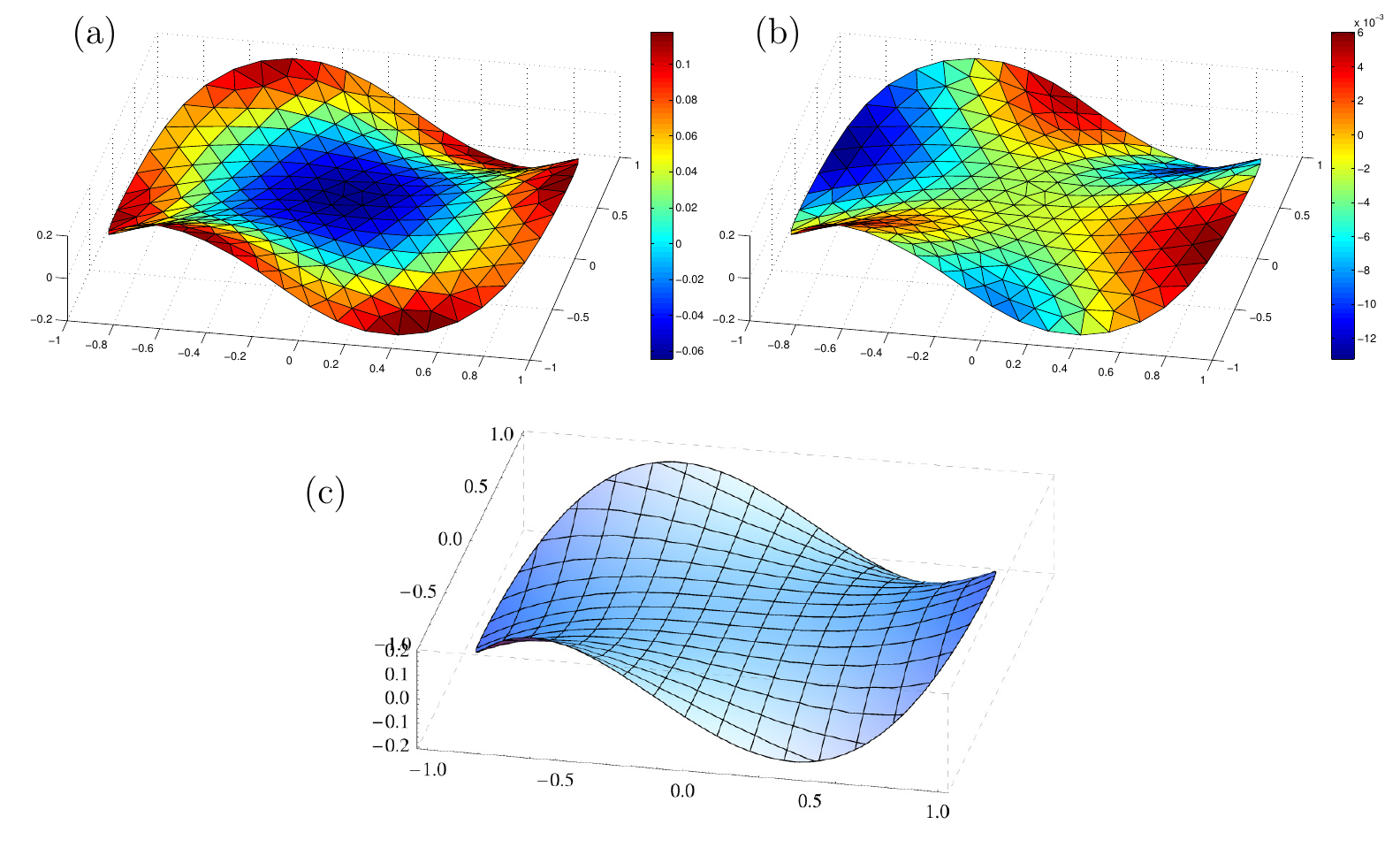}
\caption{Results of the optimization procedure for a monkey saddle target shape, showing the distribution of (a) in-plane growth $\Gamma$, and (b) active curvature change $\Psi$ in the solution. The target shape is shown in (c).}
\label{monkey}
\end{figure}

In Figure \ref{monkey}, we plot the result for a monkey saddle target shape, which has an elevation profile of $z=0.2x(x^2-3y^2)$, and $h=1$, $\nu=0.3$, $\beta=10^{-3}$, $\eta_\Gamma/\eta_\mathrm{D}=0.01$, $\eta_\Psi/\eta_\mathrm{D}=0.01$, $\eta_E/\eta_D=10$. We see clearly that the dominant factor in the solution is $\Gamma$, which is an order of magnitude greater than $\Psi$. Furthermore, $\Gamma$ is positive at the boundary of the disk but negative in the interior. This result tallies with previous results \cite{Klein-Efrati-Sharon-2007,Gemmer2013} which predict that excess growth at the boundary of the disk will cause ripples there, since the residual stress caused by the growth is relieved by buckling out of the plane.

\section{Results: one-dimensional growth}\label{sect-1d}
We can gain a greater understanding of the optimization results by considering a simplified geometry. The first example we present is of one-dimensional growth in a beam, where we set
\begin{equation}
\Gamma_{\alpha\beta}=\bra{\begin{array}{cc}\Gamma&0\\0&0\end{array}},\qquad \Psi_{\alpha\beta}=\bra{\begin{array}{cc}\Psi&0\\0&0\end{array}}
\end{equation}
and assume all quantities are independent of the Cartesian coordinate $Y$. We imposed a target shape $z=0.1\sin(\pi x)$, $0\leq x\leq1$, and considered two sets of boundary conditions. In the first case both sides are pinned:\ the displacements are fixed and a zero moment is applied. In the second case the tractions and moments at the edges are set to zero. The right-hand side is at $X=1$ because the distances have been nondimensionalized.

Due to the one-dimensional nature of the beam, $\mathcal{E}_\mathrm{D}^\mathrm{arc}$ is well-defined, and hence so is the objective function $\mathcal{E}$ in (\ref{nondim-objective}). The FvK equations (\ref{weakform1})--(\ref{weakform6}) are imposed as constraints with $h=1$. However, the simplified geometry allows us to reduce the problem to solving for $v=v_1$ and $u=w'(X)$ as piecewise affine (linear) functions over the domain, through the weak form equations
\begin{eqnarray}
\int_0^1\dd{\tilde{v}}{X}\bra{\dd{v}{X}+\frac{u^2}{2}-\Gamma}\,\mathrm{d}X=0,\\
\int_0^1\sbra{\tilde{u}u\bra{\dd{v}{X}+\frac{u^2}{2}-\Gamma}+\beta\dd{\tilde{u}}{X}\bra{\dd{u}{X}-\Psi}}\,\mathrm{d}X=0,
\end{eqnarray}
solved for all admissible variations $\tilde{v}$, $\tilde{u}$. The normal displacement is found by integrating $u$. Free boundary conditions are the natural boundary conditions while pinned boundary conditions are set by the imposition of the additional constraint that $w(1)=0$.

\begin{figure}[t]
\centering
\includegraphics[width=1\textwidth]{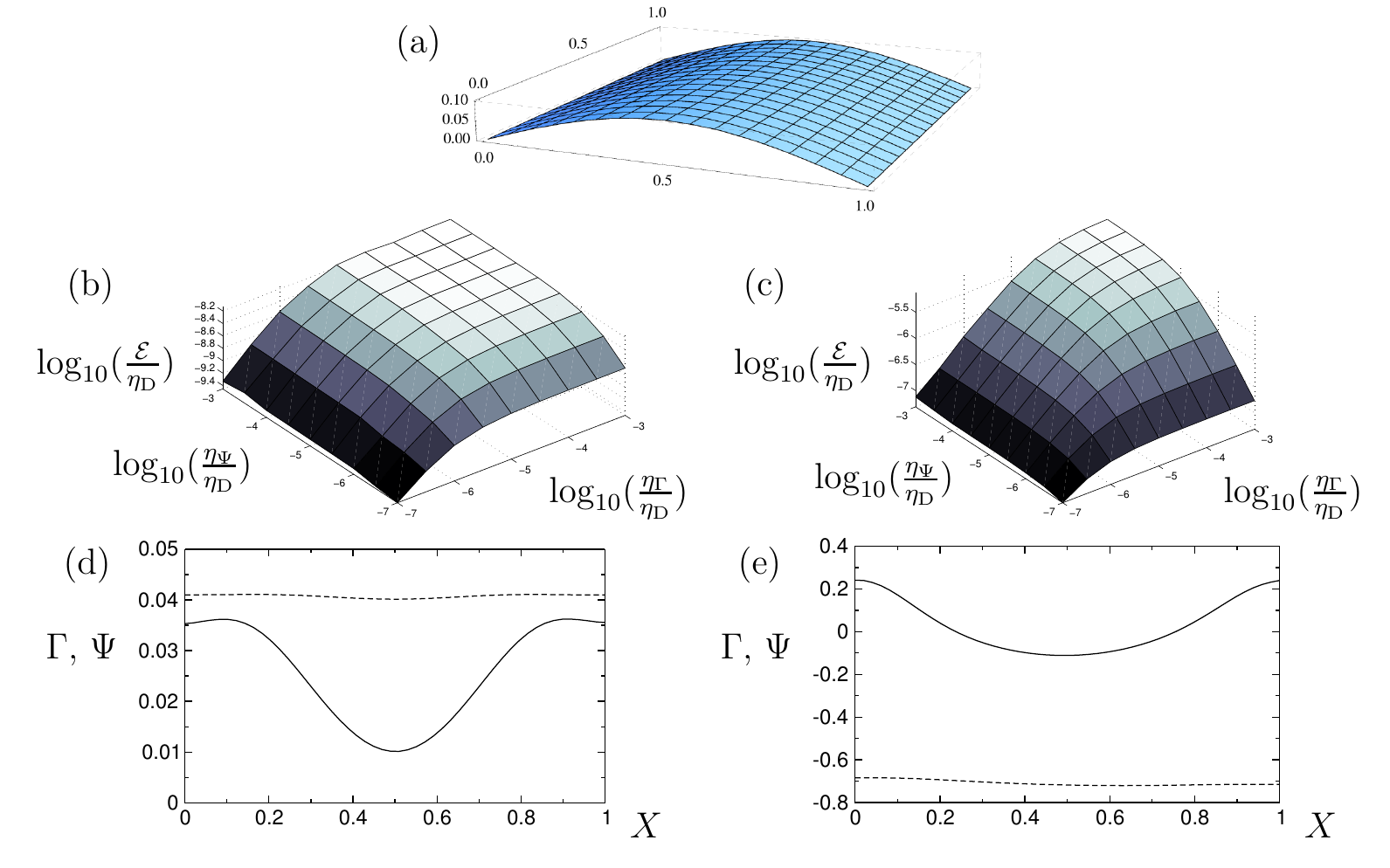}
\caption{Plot (a): Visualization of the target shape $z=0.1\sin\pi x$. Plots (b,c): Surface plots of the scaled objective function $\mathcal{E}/\eta_\mathrm{D}$ as a function of the scaled parameters $\eta_\Gamma/\eta_\mathrm{D}$ and $\eta_\Psi/\eta_\mathrm{D}$, for (b) pinned and (c) free boundary conditions. Plots (d,e): Distributions of $\Gamma$ (\full) and $\Psi$ (\dashed) over $X\in(0,1)$, for (d) pinned and (e) free boundary conditions.}
\label{1d-results}
\end{figure}

We performed sample calculations for $\beta=10^{-4}$, and for $\eta_\Gamma/\eta_\mathrm{D}$, $\eta_\Psi/\eta_\mathrm{D}$ both ranging over $10^{-7}$ to $10^{-3}$. Graphs of the objective function $\mathcal{E}/\eta_\mathrm{D}$ as a function of $\eta_\Gamma/\eta_\mathrm{D}$ and $\eta_\Psi/\eta_\mathrm{D}$ are displayed in Figure \ref{1d-results}(a,b) for both pinned and free boundary conditions. We can see that as both $\eta_\Gamma$ and $\eta_\Psi$ increase relative to $\eta_\mathrm{D}$, so does the objective function $\mathcal{E}$.

The distributions of $\Gamma$ and $\Psi$ over $X\in(0,1)$ are displayed in Figure \ref{1d-results}(c,d), for pinned and free conditions respectively. We choose the representative values of $\eta_\Gamma=10^{-6}\eta_\mathrm{D}$, $\eta_\Psi=10^{-5}\eta_\mathrm{D}$, to enforce the condition $\eta_\mathrm{D}\gg\eta_\Psi\gg\eta_\Gamma$. The reason for this choice is firstly to ensure that matching to the target shape is given the most weight, and secondly to penalize changes in $\Psi$ more than changes in $\Gamma$, since we speculate that it is simpler to experimentally control $\Gamma$ than $\Psi$.

The main difference between the solutions using different boundary conditions is that both the growth strains $\Gamma$ and $\Psi$ are larger if the edges are free. This is because in the pinned case, the plate can leverage the fixed displacements at the edges to buckle out of plane into the target shape, whereas with the free boundary condition the structure does not have this freedom (at least in one dimension) and must actively bend through $\Psi$ to achieve the shape. Indeed, solving the FvK equations directly with the calculated solution in Figure \ref{1d-results}(c) yields a bistable configuration characteristic of buckling:\ the plate can achieve both the target shape and an inverted solution, much like an Euler column (although in this case the two states have different energies, due to the asymmetry introduced through $\Psi$). This bistability is absent on using the solution in \ref{1d-results}(d).

\section{Results: axisymmetric target shapes}\label{sect-axi}
For axisymmetric target shapes, all quantities are presupposed to depend on the radial coordinate $R=\sqrt{X^2+Y^2}$ only. Beginning with a flat disk of radius $1$ (in dimensionless coordinates), we apply an isotropic growth field $\Gamma(R)$, $\Psi(R)$. Zero-traction conditions are applied on the outer rim of the disk. Given our experience of one-dimensional growth, we would thus expect $\Psi$ to play a greater role than $\Gamma$ in shaping the plate. We will also, however, repeat the calculations while holding $\Psi=0$ to see if the shapes are attainable through changes in metric alone.

\begin{figure}[t]
\centering
\includegraphics[width=1\textwidth]{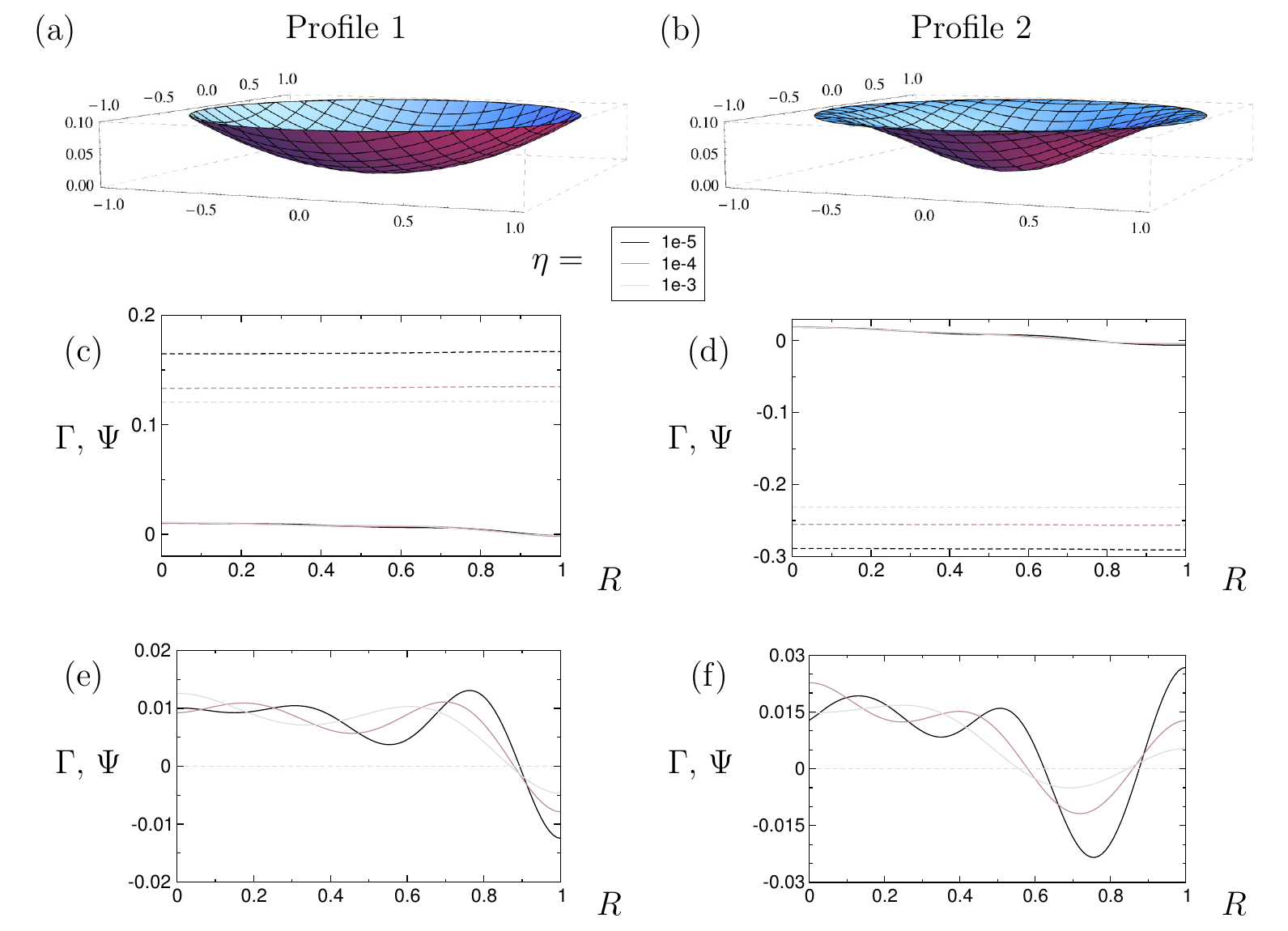}
\caption{Plots (a,b): The two target profiles used in the axisymmetric calculations. Plots (c--f): Distributions of $\Gamma$ (\full) and $\Psi$ (\dashed) for profiles 1 (c,e) and 2 (d,f), with $\eta=\eta_\Gamma/\eta_\mathrm{D}=\eta_\Psi/\eta_\mathrm{D}$. $\Psi$ is allowed to vary in plots (c,d); $\Psi$ is set to zero for plots (e,f). Inset: legend for plots (c)--(f).}
\label{axi-results}
\end{figure}

The target shape $z=f(r)$ is achieved by minimizing $\mathcal{E}_\mathrm{D}+\mathcal{E}_\mathrm{S}$ as before; the arclength functional (\ref{arclength-distance}) is used, using the cross-section of the deformed plate along the meridian $\theta=0$, without loss of generality. We perform calculations for two separate target shapes, which are displayed in Figure \ref{axi-results}(c,d):
\begin{equation}
f_1(r)=0.1r^2,\qquad f_2(r)=0.05(1-\cos\pi r),
\end{equation}
for $0\leq r\leq 1$. The Gaussian curvature of a surface defined by $z=f(r)$ can be shown to be $r^{-1}f''(r)f'(r)/(1+f'(r)^2)^2$. As such, profile 1 has a positive Gaussian curvature at all points, while the other profile consists of a central region of positive Gaussian curvature surrounded by a region of negative Gaussian curvature.

As in Section \ref{sect-1d}, there is a simplified weak form system for the solution of such axisymmetric problems. Where $u=w'(R)$, we solve the following for all admissible variations $\tilde{u}$, $\tilde{v}$:
\begin{eqnarray}
\fl\int_0^1\left[R\dd{\tilde{v}}{R}\bra{\dd{v}{R}+\frac{\nu v}{R}+\frac{u^2}{2}-(1+\nu)\Gamma}\right.\nonumber\\
\left.+\tilde{v}\bra{\nu\dd{v}{R}+\frac{v}{R}+\frac{\nu u^2}{2}-(1+\nu)\Gamma}\right]\,\mathrm{d}R=0,\\
\fl\int_0^1\left[\beta R\dd{\tilde{u}}{R}\bra{\dd{u}{R}+\frac{\nu u}{R}-(1+\nu)\Psi}+\beta\tilde{u}\bra{\nu\dd{u}{R}+\frac{u}{R}-(1+\nu)\Psi}\right.\nonumber\\
\left.+R\tilde{u}u\bra{\dd{v}{R}+\frac{\nu v}{R}+\frac{u^2}{2}-(1+\nu)\Gamma}\right]\,\mathrm{d}R=0.
\end{eqnarray}

The distributions of $\Gamma$ and $\Psi$ for the two profiles are displayed in Figure \ref{axi-results}, allowing $\Psi$ to vary (c,d) and setting it to zero (e,f). In each case, $\beta=10^{-4}$. We can clearly see that increasing $\eta =\eta_\Gamma/\eta_\mathrm{D} =\eta_\Psi/\eta_\mathrm{D}$ makes the distributions of $\Gamma$ and $\Psi$ smoother, and this is particularly noticeable when we impose $\Psi=0$. The greatest difference between the solutions with and without the assumption $\Psi=0$, is that if $\Psi\neq0$ then the solutions are almost entirely due to a constant $\Psi$ field:\ as we had predicted, the free boundary conditions mean that the plate needs to actively bend to the desired shape. It is interesting to compare the constant $\Psi$ results for both profile shapes. For the paraboloidal profile 1, the change of curvature term $\Psi$ is positive, while for profile 2 it becomes negative. We would expect the negative constant $\Psi$ to also give a paraboloidal shape, but it transpires that this state is bistable:\ a mechanical eversion gives rise to the desired profile 2. On the other hand, if $\Psi$ is set to zero, then the negative Gaussian curvature at the rim of profile 2 is introduced by increasing the growth strains here.

\section{A semi-analytic application}\label{sect-simple}
Liang and Mahadevan \cite{Liang-Mahadevan-2011} analyzed modified versions of the equations (\ref{fvk1})--(\ref{fvk2}) in order to demonstrate how a blooming flower can be regarded as a mechanical phenomenon caused by buckling due to differential growth strains. This analysis was enabled by analyzing a simplified shell geometry considered representative of the actual petal shape. Mansfield \cite{Mansfield-1962} also investigated this system --- a circular plate with zero $\Gamma_{\alpha\beta}$ and constant isotropic $\Psi_{\alpha\beta}=\Psi\delta_{\alpha\beta}$ due to an applied temperature gradient --- and showed that initially the deformed plate was a spherical cap. However, at a certain critical value of $\Psi$, this solution became unstable and bifurcated to a nonsymmetric shape similar to a section of a cylinder. This result illustrates the phenomenon of a soft mode, or a zero-stiffness deformation mode. Specifically, while the deformation field is nonaxisymmetric, the underlying mechanical properties of the material (undeformed shape, stiffness, growth fields) are independent of angle (\emph{i.e.}\ axisymmetric). Thus the same non-axisymmetric deformation, rotated by an arbitrary angle, is also a solution of the system, with the same stored energy. This one-parameter family of solutions is known as a soft mode. The ability of such structures to change shape without the requirement of large energy input has given them both theoretical and practical importance, with applications ranging from actuators to deployable structures \cite{Tarnai-2003}.

Mansfield's bifurcation was reproduced experimentally by Lee \etal \cite{Lee-Rosakis-Freund-2001}, where a flat disc comprising two layers of unequal thermal expansion coefficient was heated, corresponding to the imposition of a constant field $\Psi$ was imposed. Under a large enough temperature, the initially axisymmetric shape buckled to Mansfield's nonaxisymmetric soft mode. Other soft modes have also been developed experimentally, notably by Guest \etal \cite{Guest-Kebadze-Pellegrino-2011}, who created a zero-stiffness elastic shell by plastically deforming a metallic plate to a shell with a cylindrical geometry.

Taking Mansfield's work as our starting point, we will simplify the normal displacements and growth functions to be quadratic functions of position, and use our optimization technique to solve for the coefficients of these functions, rather than for their full pointwise distribution. We will show that the near-cylindrical geometry of Mansfield is not the only soft mode achievable by the application of axisymmetric growth functions.
These solutions are a special case of the solutions found by Seffen and Maurini \cite{Seffen-Maurini-2013}; our results emphasize the neutrally-stable nature of the deformations.

The first difficulty one encounters when performing an analysis on such a simplified deformation ansatz is that the boundary conditions will not, in general, be satisfied. To remedy this we must assume a specific form for the variable thickness. In particular, if the plate is circular, with radius $1$, set the thickness to be $h=1-R^2$. Because of the dependence of the bending and stretching stiffnesses $B$ and $D$ on $h$, we find that the in-plane stress resultants and moment resultants tend to zero as $R\to1$, so that the boundary conditions are now automatically satisfied. Additionally, with simple forms of the dependent variables, a solution may be found to the FvK equations (\ref{fvk1})--(\ref{fvk2}). For instance, for a circular plate of (dimensionless) radius $1$, set
\begin{eqnarray}
h=1-x^2-y^2=1-R^2,\label{ansatz1}\\
\chi=kh^3,\qquad\qquad\Gamma=\Gamma_0+\Gamma_2R^2,\qquad\qquad\Psi=\Psi_0+\Psi_2R^2,\\
w=\kappa_1X^2+\kappa_2Y^2=R^2(\kappa_1\cos^2\theta+\kappa_2\sin^2\theta),\label{ansatz3}
\end{eqnarray}
where we have assumed isotropic growth. We have hereby reduced the problem to determining the seven constants $\Gamma_0$, $\Gamma_2$, $\Psi_0$, $\Psi_2$, $k$, $\kappa_1$, and $\kappa_2$ by minimizing the objective function $\mathcal{E}$ subject to the FvK equation constraints. Considering the constraints first, the stress-free boundary conditions for this system are satisfied automatically. On substituting (\ref{ansatz1})--(\ref{ansatz3}) into the FvK equations (\ref{fvk1})--(\ref{fvk2}), we obtain the following relations between the coefficients:
\begin{eqnarray}
6k(7+\nu)+\Gamma_2(1-\nu^2)+\kappa_1\kappa_2(1-\nu^2)=0,\label{semianalytic-constr-1}\\
(\kappa_2-\kappa_1)(k+\beta(1-\nu))=0,\label{semianalytic-constr-2}\\
\beta(1+\nu)\Psi_0+(\kappa_1+\kappa_2)(k-\beta(1+\nu))=0,\label{semianalytic-constr-3}\\
\Psi_2=0.\label{semianalytic-constr-4}
\end{eqnarray}

The remaining three degrees of freedom are set by minimizing the objective function $\mathcal{E}$. Since $\Psi_2=0$, $\boldsymbol{\nabla}\Psi=0$ and hence $\mathcal{E}_\mathrm{S}=\eta_\Gamma\iint|\boldsymbol{\nabla}\Gamma|^2\mathrm{d}^2\boldsymbol{X}=2\pi\eta_\Gamma\Gamma_2^2$. However, for this application the smoothness of $\Gamma$ is not relevant and we set $\eta_\Gamma=0$, so that $\mathcal{E}=\mathcal{E}_\mathrm{D}$.

To calculate $\mathcal{E}_\mathrm{D}$, we need the full displacement field, including the in-plane displacements $v_R$ in the radial direction and $v_\theta$ in the circumferential direction. These are given by
\begin{eqnarray}
\fl v_R(R,\theta)=R\Gamma-\frac{R^3(\kappa_1-\kappa_2)^2(3+\cos4\theta)}{12}\nonumber\\
+\frac{R^3(\kappa_2^2-\kappa_1^2)\cos2\theta}{3}+\frac{6kR(\nu(1-R^2)+5R^2-1)}{1-\nu^2},\\
\fl v_\theta(R,\theta)=\frac{R^3(\kappa_1^2-\kappa_2^2)\sin2\theta}{6}+\frac{R^3(\kappa_1-\kappa_2)^2\sin4\theta}{12}.
\end{eqnarray}
The expression for $v_\theta$ does not lend itself well to a simple distance function which may be integrated over the area of the circular plate. However, we may approximate a distance function by calculating the arclength distance measure (\ref{arclength-distance}) for $\theta=0,\frac{\pi}{2},\pi,\frac{3\pi}{2}$ (which is where $v_\theta=0$), and summing the results:
\begin{equation}
\mathcal{E}_D=\mathcal{E}_D^\mathrm{arc}|_{\theta=0}+\mathcal{E}_D^\mathrm{arc}|_{\theta=\pi/2}+\mathcal{E}_D^\mathrm{arc}|_{\theta=\pi}+\mathcal{E}_D^\mathrm{arc}|_{\theta=3\pi/2}.
\end{equation}

We can now state the optimization problem for this simplified formulation (F1): Choose $\Gamma_0$, $\Gamma_2$, $k$, $\kappa_1$, $\kappa_2$, $\Psi_0$ that minimize $\mathcal{E}_\mathrm{D}$, subject to equations (\ref{semianalytic-constr-1})--(\ref{semianalytic-constr-3}).

We will now illustrate this method by considering the growth patterns required to transform the circular plate to the targets outlined earlier. We consider an axisymmetric profile $z=\lambda(x^2+y^2)$, a cylindrical profile $z=\lambda x^2$ (as an approximation to Mansfield's bifurcated solution) and a saddle geometry $z=\lambda(x^2-y^2)$.

Substituting these target shapes into the optimization procedure will output the values of the constants. However, by exploiting symmetry to write $\kappa_2$ in terms of $\kappa_1$, we can find $\Psi_0$, $\Gamma_2$ and $k$ in terms of $\kappa_1$ from (\ref{semianalytic-constr-1})--(\ref{semianalytic-constr-3}), and then $\Gamma_0$ and $\kappa_1$ are calculated by minimizing the distance functional $\mathcal{E}_\mathrm{D}$.

\begin{figure}[ht]
\centering
\includegraphics[width=0.6\textwidth]{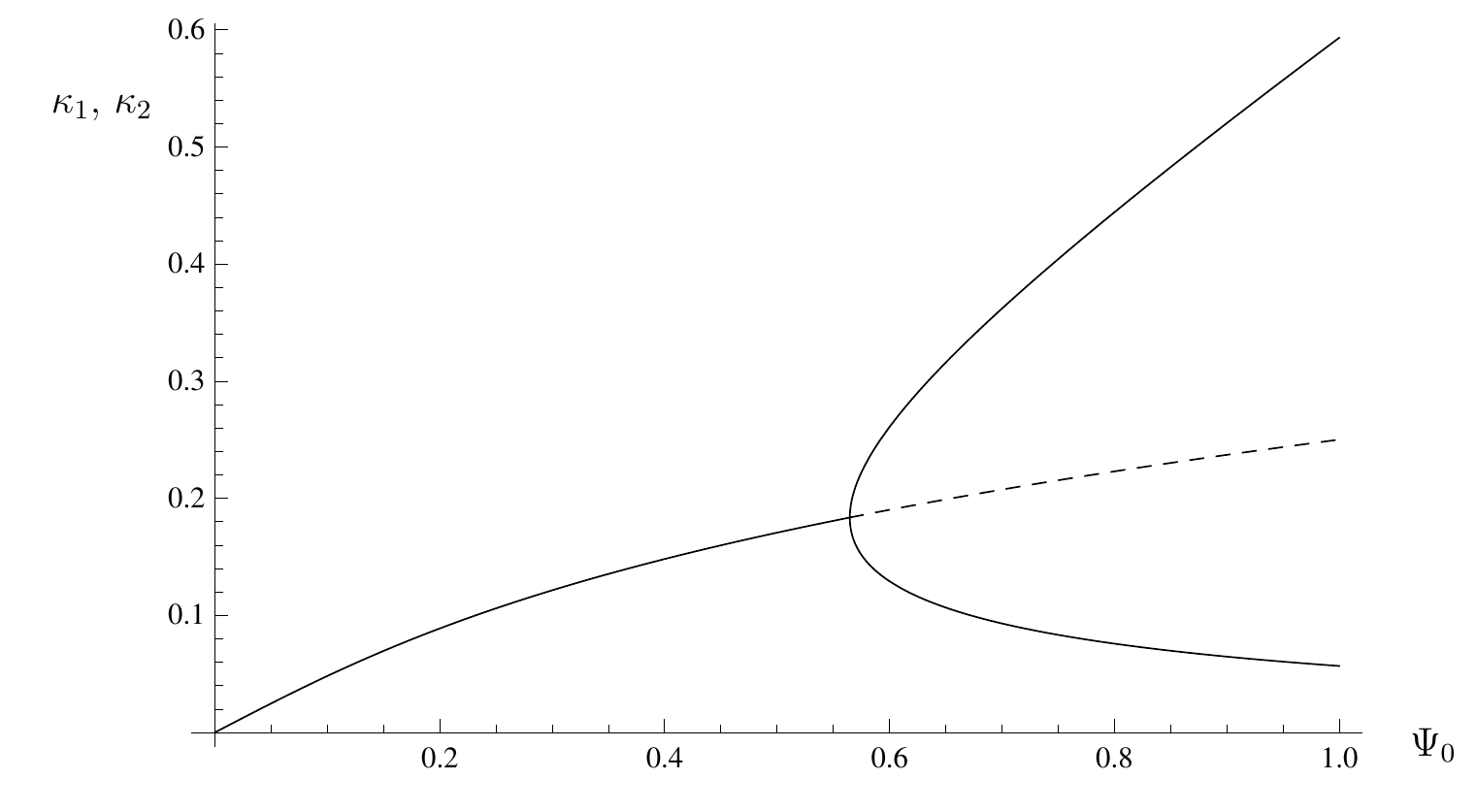}
\caption{The bifurcation problem described by Mansfield \cite{Mansfield-1962}. As $\Psi_0$ increases past a critical value, the axisymmetric solution ($\kappa_1=\kappa_2$) becomes unstable, and a solution with $\kappa_1\neq\kappa_2$ emerges. Plots are for $\nu=0.3$ and $\beta=10^{-3}$.}
\label{mansfield-bifurcation}
\end{figure}

For the paraboloid of revolution $z=\lambda(x^2+y^2)$, $\kappa_2=\kappa_1$ and equation (\ref{semianalytic-constr-2}) is automatically satisfied. The optimization thus has an extra degree of freedom. However, the solution obtained has $\Psi_0$ much greater than both $\Gamma_0$ and $\Gamma_2$, making it comparable with Mansfield's original solution with $\Gamma=0$. In fact, setting $\Gamma_2=0$ we obtain his result exactly:
\begin{equation}
\Psi_0=2\kappa_1+\frac{\kappa_1^3(1-\nu)}{3\beta(7+\nu)}.
\end{equation}
However this result, as noted previously, becomes unstable when $\Psi_0^2>96\beta(7+\nu)/(1+\nu)^3$. Figure \ref{mansfield-bifurcation} displays the bifurcation diagram for the parameters $\kappa_1$, $\kappa_2$ as $\Psi_0$ varies.

For those cases where the $\Psi_0$-only solution is unstable, we can still find a paraboloidal solution by setting $\Psi_0=0$; by subsequently solving (\ref{semianalytic-constr-1})--(\ref{semianalytic-constr-3}) we obtain $\Gamma_2$. In summary:
\begin{equation}
\Gamma_2=-\kappa_1^2-\frac{6\beta(7+\nu)}{1-\nu},\qquad \Psi_0=0.\label{G2-P0-bowl}
\end{equation}
In both cases, $\Gamma_0$ and $\kappa_1$ are found by minimizing $\mathcal{E}_\mathrm{D}$. Note the similarity between these results and those of section \ref{sect-axi}, where a paraboloidal bowl was found for $\Psi=\textrm{constant}$, or for $\Psi=0$ and an in-plane growth which may be approximated by $\Gamma=\Gamma_0-\Gamma_2R^2$, as here.

For a cylindrical target shape, $\kappa_2=0$ (as opposed to Mansfield's bifurcated solution, which had both $\kappa_1$ and $\kappa_2$ positive). We can achieve this shape by solving (\ref{semianalytic-constr-1})--(\ref{semianalytic-constr-3}) to give
\begin{equation}
\Gamma_2=6\beta(7+\nu)/(1+\nu),\qquad\Psi_0=2\kappa_1/(1+\nu),\label{G2-P0-fold}
\end{equation}
with $\Gamma_0$ and $\kappa_1$ again solved for by optimization of the distance functional.

\begin{figure}[t]
\centering
\includegraphics[width=0.8\textwidth]{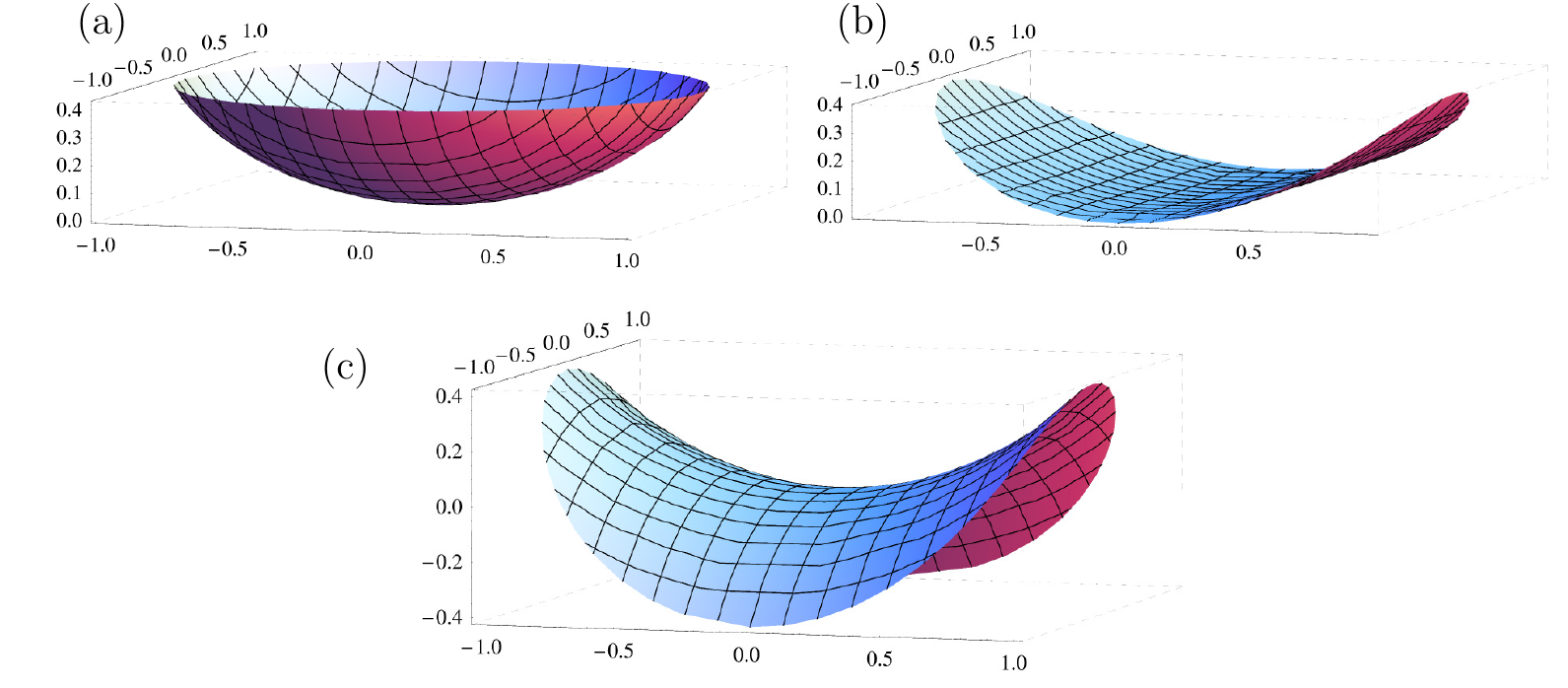}
\caption{Plots of the deformed plates under growth strains $\Gamma=\Gamma_0+\Gamma_2R^2$, $\Psi=\Psi_0$, where the coefficients are chosen to satisfy the minimization problem (F1). The target shape $z=f(x,y)$ and specific expressions for $\Gamma_2$, $\Psi_0$ are given as follows: (a) Paraboloidal target, $f(x,y)=\lambda(x^2+y^2)$, equations (\ref{G2-P0-bowl}); (b) Cylindrical target, $f(x,y)=\lambda x^2$, equations (\ref{G2-P0-fold}); (c) Saddle-shaped target, $f(x,y)=\lambda(x^2-y^2)$, equations (\ref{G2-P0-pringle}).}
\label{sa-plots}
\end{figure}

Finally a saddle shape where $\kappa_2=-\kappa_1$ can be found in the same manner:\ this time
\begin{equation}
\Gamma_2=\kappa_1^2+6\beta(7+\nu)/(1+\nu),\qquad\Psi_0=0.\label{G2-P0-pringle}
\end{equation}
Plots of all three deformed plates can be seen in Figure \ref{sa-plots}.

\section{Conclusions and extensions}
In this article we have outlined a new approach to determining the optimal distribution of growth stresses that transform a flat plate into a specified target shape. Not only have we calculated the solution for non-symmetric and for simplified one-dimensional geometries (Sections \ref{sect-2d}--\ref{sect-axi}), but qualitative results have been obtained using a semi-analytic approach (Section \ref{sect-simple}), and have been used to show that an axisymmetric growth pattern can be used to produce a structure which exhibits soft mode deformations.

Possible extensions to this theory include curved initial geometries (shells), the relaxation of the small-growth-strain assumptions (leading to more strongly nonlinear equations), and the use of different control variables, such as edge displacements or surface tractions. In any case we believe that this approach will prove useful for researchers who wish to engineer plate deformations into a desired shape.

\section*{Acknowledgements}
The authors would like to acknowledge funding from the Harvard--National Science Foundation Materials Research Science and Engineering Center, the Wyss Institute for Biologically Inspired Engineering, and the Kavli Institute for Bionano Science and Technology.

\appendix

\section{Two-dimensional growth optimization solution procedure}\label{proc-appendix}
Here we outline the solution procedure for the problem described in section \ref{sect-2d}. For this, the equations (\ref{weakform1})--(\ref{weakform6}) require discretization. The state variables $w$, $v_1$, $v_2$, $\rho_{11}$, $\rho_{12}$, $\rho_{22}$ and control variables $\Gamma$, $\Psi$ are defined in terms of their values at $N$ points forming the nodes of a triangulation of the domain $\Omega$. The triangulation enables the generation of basis functions $\phi_i(X,Y)$ for each node $i=1,\ldots,N$, so that (for instance) the out-of-plane displacement is approximated by $w\approx\sum_{i=1}^Nw_i\phi_i$.

This allows the six weak form PDEs (\ref{weakform1})--(\ref{weakform6}) to be rewritten as $6N$ algebraic equations in terms of the nodal values of the variables.

\medskip

\noindent {\bf Computational procedure:}
\begin{enumerate}
\item Express the outline of the initial ungrown plate as a parametric representation $(X_\mathrm{b}(\theta),Y_\mathrm{b}(\theta))$.
\item Express a target surface $g(x,y,z)=z-f(x,y)$ for the grown plate, together with a target boundary $(x_\mathrm{b}(\theta),y_\mathrm{b}(\theta),z_\mathrm{b}(\theta))$.
\item Calculate $(x_\mathrm{b}(\theta_s(s_\mathrm{max}\sigma)), y_\mathrm{b}(\theta_s(s_\mathrm{max}\sigma)), z_\mathrm{b}(\theta_s(s_\mathrm{max}\sigma)))$  from (\ref{xbybzb}) for a fine mesh of $\sigma\in[0,1]$.
\item Use $(X_\mathrm{b}(\theta), Y_\mathrm{b}(\theta))$ to find a triangulation of the source domain $\Omega$.
\item For each node $i$ in the triangulation, calculate the basis functions $\phi_i(X,Y)$.
\item Initialize the state and control variables to be zero at each node.
\item {\bf Main solution routine.} The optimization routine \verb|e04vh| calculates the optimal $y_j$ ($j=1,\ldots,m$) such that $F_1(y_j)$ is minimized subject to $F_i^{\mathrm{min}}\leq F_i(y_j)\leq F_i^{\mathrm{max}}$ for $i=2,\ldots,n$. In our case the number of equations $n-1$ is $6N$, and the number of variables $m$ is $8N$.
\begin{enumerate}
\item {\bf Limits:} Set $F_i^{\mathrm{min}}=0$ for each $i=1,\ldots,6N+1$, $F_1^{\mathrm{max}}=\infty$, and $F_i^{\mathrm{max}}=0$ for each $i=2,\ldots,6N+1$.
\item {\bf Subroutine:} calculate $F_i$ given $y_j$.
\begin{enumerate}
\item The input vector $y_j$ is the concatenation of the values of $v_1$, $v_2$, $w$, $\rho_{11}$, $\rho_{12}$, $\rho_{22}$, $\Gamma$, and $\Psi$ at each node in the triangulation.
\item Use the triangulation geometry and basis functions $\phi_i$ to calculate the gradients of each of these variables in each triangle (by construction, they will be piecewise constant in each triangle).
\item Calculate the value of $\mathcal{D}^2=g(X+v_1,Y+v_2,w)^2$ at each node in the triangulation, and use this to calculate $\eta_\mathrm{D}\iint_\Omega\mathcal{D}^2\,\mathrm{d}^2\boldsymbol{X}$.
\item Find the boundary of the deformed mesh, and calculate
\begin{equation}
(x_\mathrm{b}^\mathrm{d}(\theta_S(S_\mathrm{max}\sigma_k)), y_\mathrm{b}^\mathrm{d}(\theta_S(S_\mathrm{max}\sigma_k)), z_\mathrm{b}^\mathrm{d}(\theta_S(S_\mathrm{max}\sigma_k)))
\end{equation}
for each point $k$ corresponding to a boundary node. Use this together with the previously calculated
\begin{equation}
(x_\mathrm{b}(\theta_s(s_\mathrm{max}\sigma)), y_\mathrm{b}(\theta_s(s_\mathrm{max}\sigma)), z_\mathrm{b}(\theta_s(s_\mathrm{max}\sigma)))
\end{equation}
for these $\sigma_k$ to calculate $\mathcal{E}_\mathrm{E}$ from (\ref{EE}).
\item Use the gradients of $\Gamma$ and $\Psi$ to calculate $\mathcal{E}_\mathrm{S}$.
\item Combine the previous three integrals to calculate the objective function, and set $F_1$ to be this value.
\item Calculate the $6N$ discretized weak form equations, and set these to be the constraints $F_2,\ldots,F_{6N+1}$.
\end{enumerate}
\end{enumerate}
\item Output state and control variables and plot results.
\end{enumerate}

\section*{References}

\providecommand{\newblock}{}

\end{document}